\documentclass[runningheads]{llncs}

\usepackage{graphicx}
\usepackage{comment}
\usepackage{amsmath,amssymb}
\usepackage{color}
\usepackage{url}
\usepackage{hyperref}
\usepackage{graphicx,verbatim}
\usepackage{booktabs}

\newif\ifreview

\reviewfalse

\ifreview
	\usepackage{lineno}

	\linenumbers
\fi

\begin{document}

\def\SubNumber{31}
\def\GCPRTrack{Main Track}

\title{Enhancing Synthetic CT from CBCT via Multimodal Fusion: A Study on the Impact of CBCT Quality and Alignment}
\titlerunning{Enhancing Synthetic CT via Multimodal Fusion}

\ifreview
\else

	\author{Maximilian Tschuchnig\inst{1,3}\orcidID{0000-0002-1441-4752} \and
	Lukas Lamminger\inst{2} \and
    Philipp Steiningerr\inst{2} \and
	Michael Gadermayr\inst{1}\orcidID{0000-0003-1450-9222}}
	
	\authorrunning{M. Tschuchnig et al.}
	
	\institute{Salzburg University of Applied Sciences, Puch bei Hallein, Austria \and MedPhoton GmbH, Salzburg, Austria \and University of Salzburg, Salzburg, Austria
	\email{maximilian.tschuchnig@fh-salzburg.ac.at}}
\fi


%

%
\maketitle              
\begin{abstract}
Cone-Beam Computed Tomography (CBCT) is widely used for real-time intraoperative imaging due to its low radiation dose and high acquisition speed. However, despite its high resolution, CBCT suffers from significant artifacts and thereby lower visual quality, compared to conventional Computed Tomography (CT). A recent approach to mitigate these artifacts is synthetic CT (sCT) generation, translating CBCT volumes into the CT domain. In this work, we enhance sCT generation through multimodal learning, integrating intraoperative CBCT with preoperative CT. Beyond validation on two real-world datasets, we use a versatile synthetic dataset, to analyze how CBCT-CT alignment and CBCT quality affect sCT quality. The results demonstrate that multimodal sCT consistently outperform unimodal baselines, with the most significant gains observed in well-aligned, low-quality CBCT-CT cases. Finally, we demonstrate that these findings are highly reproducible in real-world clinical datasets.


\end{abstract}
\section{Introduction}
\sloppy Mobile robotic imaging systems, such as cone-beam computed tomography (CBCT)~\cite{rafferty2006intraoperative}, provide real-time, intraoperative imaging, facilitating guidance during medical procedures, and is especially useful in radiation therapy~\cite{thummerer2023synthrad}. CBCT, a specific type of computed tomography (CT), uses a cone-shaped X-ray beam to acquire 3D images in a single rotation, reducing acquisition time and radiation exposure. However, CBCT images typically suffer from lower contrast-to-noise ratios and increased artifacts compared to CT scans~\cite{wei2024reduction}.

One approach to mitigating CBCT artifacts, as highlighted by Altalib et al.~\cite{altalib2025synthetic}, is the conversion of high-artifact CBCT volumes into a low-artifact CT domain using image-to-image translation. The resulting, so-called synthetic CTs (sCTs)~\cite{altalib2025synthetic,chen2020synthetic} aim to combine specific domain advantages, here the speed and mobility advantages of CBCT and improved image quality (less artifacts) of standard CT scans. The importance of robust sCT generation was underscored by the SynthRad Grand Challenge 2023~\cite{thummerer2023synthrad}, which exposed key limitations of existing methods, especially their sensitivity to CBCT artifacts and anatomical misalignment, and motivated the need for more resilient techniques.

However, sCTs face two key limitations. First, they cannot create new anatomical details beyond those present in the input data, and the effectiveness of training models generating sCT depends on the availability of high-quality training data. While the latter can be mitigated through the use of data augmentation and synthetic data, the former requires integrating additional information, such as high-quality preoperative CT scans, into the sCT generation process~\cite{altalib2025synthetic,chen2021synthetic,chen2020synthetic}.

High-quality preoperative CT scans are typically acquired before a procedure to aid in treatment planning, whereas intraoperative CBCT scans are obtained immediately before or during the procedure for real-time organ or tumor localization. For instance, in radiotherapy, treatment plans are initially based on high-resolution preoperative CT scans, while intraoperative CBCT is employed to refine tumor and organ-at-risk localization before radiation delivery~\cite{hong2022ct}.  

While sCT methods have been shown to enhance visual image quality~\cite{altalib2025synthetic}, we hypothesize that their accuracy can be further improved by integrating intraoperative CBCT with preoperative CT. This is motivated by the complementary strengths of both modalities: CBCT provides real-time anatomical information during the procedure, while CT offers high-quality structural detail acquired during planning. Multimodal learning, which leverages such complementary information, has shown potential in sCT generation~\cite{chen2020synthetic,chen2021synthetic}. Recent advances in multi-modality fusion from computer vision, such as the self-supervised Equivariant Multi-Modality Image Fusion framework for heterogeneous data integration~\cite{zhao2024emma}, further demonstrate the potential to effectively combine multiple imaging sources. Various fusion strategies exist~\cite{zhang2021deep,zhang2021modality}, including early, late, and hybrid fusion. In this study, we focus on early fusion—the most common approach—where images from multiple modalities are combined before being processed by an sCT generation model.

Our contributions are threefold: (1) We extend existing 2D multimodal synthetic CT (sCT) generation approaches to a 3D framework, enabling improved spatial consistency and leveraging preoperative CT as complementary anatomical information. (2) We systematically evaluate, using a synthetic dataset, how CBCT quality and CBCT–CT alignment impact sCT generation performance, showing that multimodal learning consistently outperforms unimodal baselines, particularly in well-aligned, low-quality CBCT scenarios. (3) We validate the robustness and generalizability of our approach on two additional real-world, open-access datasets, demonstrating its effectiveness across diverse clinical acquisition settings.

\section{Methodology and Materials}
We hypothesize that by adding high-quality, preoperative planning CT to intraoperative CBCT with a comparably high degree of CBCT specific artifacts, sCT quality can be improved, even if CBCT and CT are imperfectly aligned. This approach is inspired by the 2D model introduced by Chen et. al.~\cite{chen2021synthetic}, which combines intraoperative CBCT with preoperative planning CT for sCT generation. Similarly, Tschuchnig et. al.~\cite{tschuchnig2024initial} used 3D multimodal learning of preoperative CT and intraoperative CBCT for organ and tumor segmentation. Both approaches demonstrate that fusing intraoperative CBCT with planning CT can enhance downstream performance. In addition, we investigate how intraoperative image quality and CBCT-CT alignment influence unimodal and multimodal sCT generation. To facilitate reproducibility, all code used for these experiments is publicly available on GitHub at: [Blinded].

\subsection{3D Multimodal sCT Generation}
The proposed 3D multimodal sCT generation method combines deep learning with multimodal learning by adapting a 3D image reconstruction model, specifically 3D U-Net~\cite{cciccek20163d}, with early fusion multimodal learning~\cite{podobnik2023multimodal}. 
The model $F$ takes as input a combination of preoperative, unaligned CT volumes $V_{CT}$ and intraoperative CBCT volumes $V_{CBCT}$. These volumes are concatenated along a new fourth dimension, analogous to how RGB channels are treated in 2D images, to form the combined volume $V=[V_{CT},V_{CBCT}]$. This volume is then fed into $F$ to generate the corresponding sCT $\hat{Y}$. Using $\hat{Y}$ and the preoperative, aligned CT $Y$ the loss is calculated with $L(\hat{Y},Y)$. As baselines, unimodal sCT models are applied using only CBCT as input. To assess whether the multimodal model primarily relies on the unaligned CT while disregarding the intraoperative CBCT, we introduce an additional baseline (CT-only), which uses only the unaligned CT as input and compares the output to the aligned ground truth CT. However, this evaluation requires access to a true sCT ground truth, available only in the synthetic dataset, making it inapplicable to real-world datasets.

\begin{figure}
\includegraphics[width=\textwidth]{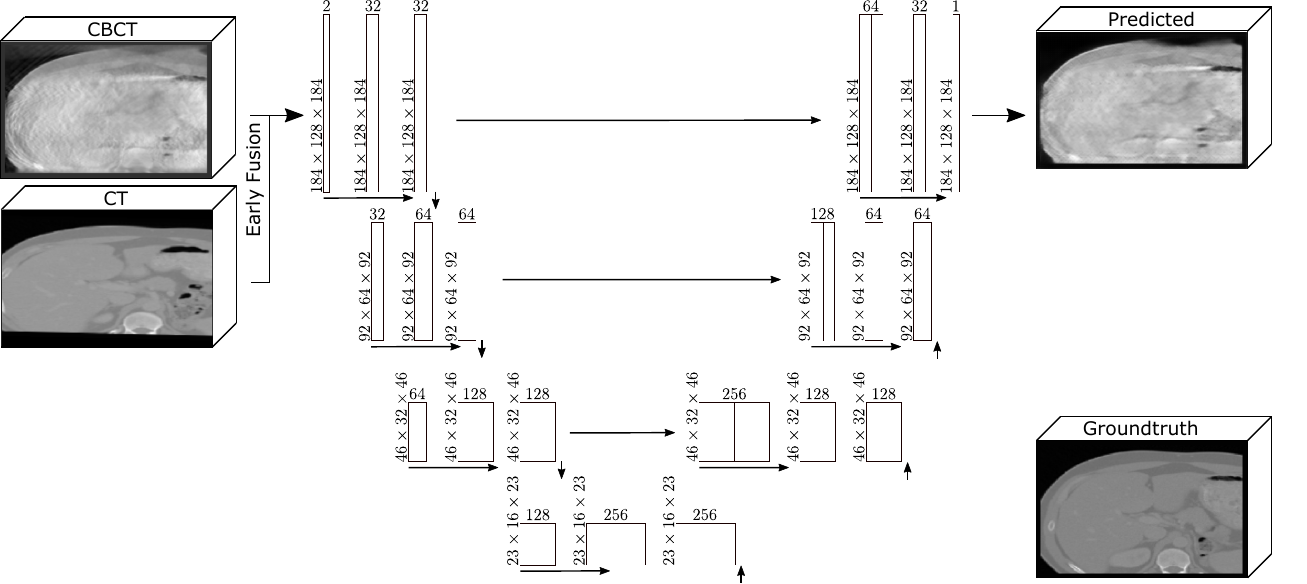}
\caption{Proposed 3D, (early fusion) multimodal sCT model, based on U-Net.} \label{Fig::I2IMethod}
\end{figure}

The proposed multimodal, 3D sCT generation model is based on 3D U-Net, introduced by Çiçek et al.~\cite{cciccek20163d} and shown in Fig.~\ref{Fig::I2IMethod}. In detail, it consists of an encoder with 3 double convolution layers with $3 \times 3 \times 3$ convolutional kernels, connected by 3D max pooling. The latent space consists of one double convolution block followed by the U-Net decoder, mirroring the encoder. Each double convolutional output in the encoder is also connected to the decoder double convolutional block of the same order. Additionally, one 3D convolutional layer is added to the decoder with a filter size of $1 \times 1 \times 1$ and the number of filters set to $1$ to facilitate image reconstruction. No transfer function is applied to the last layer, returning logits directly. The number of feature maps is set to $\{32, 64, 128, 256\}$. Batch norm is applied after each layer in the double convolutional blocks. For multimodal early fusion, the model input is composed of two feature maps, as opposed to the single feature map input in the original 3D U-Net~\cite{cciccek20163d}.

To train the model we use a weighted sum of voxel, patch based and perceptual losses, as typically used in image reconstruction and image-to-image based models~\cite{johnson2016perceptual,chen2020synthetic,altalib2025synthetic}. For a voxel based loss we apply mean absolute error (MAE) to focus on preserving image structure and robustness to noise. As a patch based loss, the structured similarity index measure (SSIM) is applied to also focus on the patch information luminance, contrast and structure. To convert the similarity into a loss function, we utilize $1-SSIM(\hat{Y},Y)$.
Additionally, a perceptual loss~\cite{johnson2016perceptual} based on vgg16~\cite{Simonyan2014VeryDC}, pretrained on the imagenet dataset~\cite{deng2009imagenet} is added to preserve perceptual information. In detail, we use the pretrained vgg16 to extract mid-level feature maps (vgg16 up to and including layer $23$) from the sCT and compare these feature maps to the original CTs vgg16 feature maps using MAE. Therefore, our perceptual loss is defined as $MAE(vgg_{:23}(\hat{Y}),vgg_{:23}(Y))$. Our final loss is given by $\alpha_1 \cdot MAE(\hat{Y},Y) + \alpha_2 \cdot 1-SSIM(\hat{Y},Y) + \alpha_3 \cdot MAE(vgg_{:23}(\hat{Y}),vgg_{:23}(Y))$ with $\alpha_1=0.2$, $\alpha_2=0.1$ and $\alpha_3=0.7$. Loss weights were empirically chosen based on validation metrics and qualitative assessment. Perceptual loss was prioritized to improve sharpness and anatomical detail. In qualitative observations, relying solely on perceptual loss led to hallucinated structures, which were mitigated by incorporating the voxel-wise MAE term. SSIM further complemented both by encouraging local structural coherence and contrast preservation.

\subsection{Dataset}
To thoroughly investigate the effects of intraoperative image quality and alignment on sCT generation performance, a dataset is utilized which includes varying levels of intraoperative CT image quality with perfect alignment between preoperative CT and intraoperative CBCT. 
While perfect alignment between preoperative CT and intraoperative CBCT is rarely achievable in real clinical settings, synthetically aligned data plays a valuable role during model development. It allows for controlled experiments to systematically study the effects of alignment and image quality on sCT generation. Nonetheless, real-world datasets remain essential for evaluating the model’s practical robustness and generalizability under realistic clinical conditions, where misalignments and acquisition variability naturally occur.
Importantly, while perfect CBCT-CT alignment is not realistic during inference, synthetically aligned data remains valuable for model development and training. Evaluating the proposed multimodal learning approach is particularly challenging when using real, clinical data.
In these scenarios, CBCT-CT alignment is often imperfect (even if performed manually or semi-automatically) due to patient movement and respiratory variations. Even with controlled breathing techniques designed to minimize motion-induced misalignment, perfect CBCT-CT alignment remains difficult to achieve~\cite{hong2022ct}. 
Furthermore, assessing the impact of CBCT visual quality using real patient data is problematic, as clinical practice aims to optimize CBCT acquisitions. Deliberately acquiring lower-quality CBCT scans would expose patients to unnecessary discomfort and radiation and is not standard practice. Instead, synthetic CBCT volumes~\cite{tschuchnig2024cbctlits}, generated from CT scans, offer a viable alternative by ensuring perfect alignment and enabling controlled variations in CBCT quality through undersampling techniques.  

\begin{figure}[t]
\includegraphics[width=\textwidth]{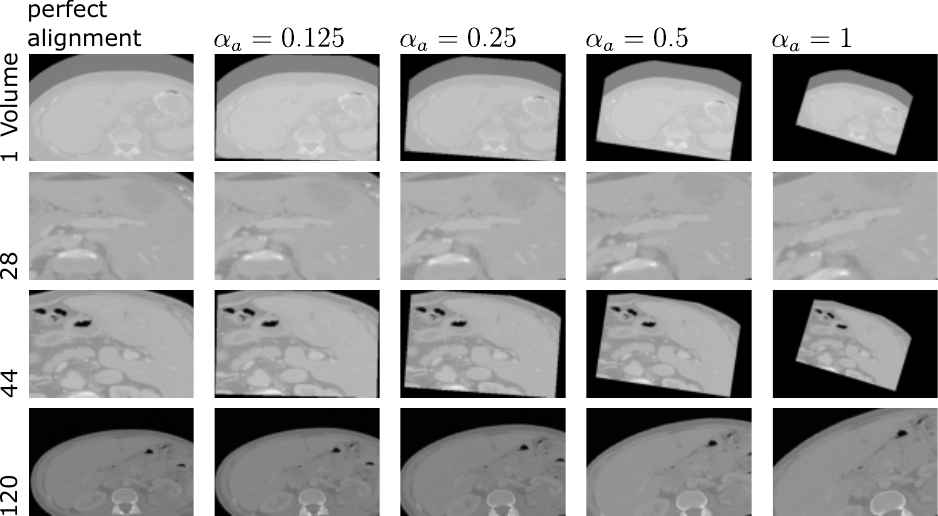}
\caption{Sample results, showing the original CT as well as synthetically unaligned versions of the same CT (CBCTLiTS dataset). Four different volumes are shown with increasing $\alpha_a$ (unalignment), displaying rotation, scaling and minor translation.} \label{Fig::DataAugResults}
\end{figure}

We utilize three datasets.
1: CBCT Liver Tumor Segmentation Benchmark (CBCTLiTS) dataset~\cite{tschuchnig2024cbctlits}. CBCTLiTS consists of $131$ synthetic CBCT and corresponding real CT images of the abdominal region. The paired volumes are perfectly aligned and available in five different quality levels, enabling a controlled study of intraoperative image quality ($\alpha_{np}$) and alignment ($\alpha_a$).  
2: Pancreatic-CT-CBCT-SEG dataset~\cite{hong2022ct} (pancreas dataset) The pancreas dataset provides $40$ CBCT-CT pairs of the abdominal region, serving as a challenging benchmark to validate our findings in a clinical setting. Unlike CBCTLiTS, the pancreas dataset includes naturally occurring misalignment and relatively uniform CBCT quality.  
3: SynthRAD2023~\cite{thummerer2023synthrad} dataset (synthrad). SynthRAD provides 360 paired CT-CBCT and CT-MR volumes across brain and pelvis anatomies, acquired from three clinical centers. The dataset includes both rigidly registered CBCT and MRI images to planning CTs, enabling sCT generation tasks across varying acquisition protocols and image qualities. For our experiments, we chose the target of pelvis sCT generation from CBCT.

All datasets used are publicly available: CBCTLiTS under CC BY-NC-ND 4.0, Pancreatic-CT-CBCT-SEG under CC BY 4.0, and SynthRAD under CC BY-NC 4.0.

To assess the impact of alignment on multimodal sCT reconstruction, the parameter $\alpha_a$ is introduced. This factor, defined as $\alpha_a = [0,1]$, describes how strongly the preoperative CT is artificially unaligned. $\alpha_a$ reduces the number of parameters controlling unalignment by combining multiple affine unalignments (rotation, scaling and translation) into a single parameter. Misalignment was performed using TorchIO RandomAffine. In detail, affine misaligned was performed using random (non-isotropic) scaling, with the scaling parameter sampled from $\mathcal{U}(1 - 0.5 \cdot \alpha_a,1 + 0.5 \cdot \alpha_a)$, rotation, parameters sampled from $\mathcal{U}(-22.5 \cdot \alpha_a, 22.5 \cdot \alpha_a)$, and translation, with the parameter sampled from $\mathcal{U}(0, 0.05 \cdot \alpha_a)$ with tri-linear interpolation. Sample results of this unalignment on the CBCTLiTS dataset are shown in Fig.~\ref{Fig::DataAugResults}. All datasets were unaligned in the same way.

\sloppy To further investigate $\alpha_a$ unalignment, the average voxel distance between the original and unaligned volumes was calculated using the perfectly paired CBCTLiTS dataset. The resulting average voxel-wise displacement (in voxels) for different $\alpha_a$ values were ${(\alpha_a=0:0),(\alpha_a=0.125:2.6),(\alpha_a=0.25:3.9),(\alpha_a=0.5:6),(\alpha_a=1:7.8)}$, corresponding to physical distances of approximately ${0\mathrm{mm}, 2.1\mathrm{mm}, 3.1\mathrm{mm}, 4.8\mathrm{mm} and 6.3\mathrm{mm}}$, respectively. These values cover a realistic range from mild misalignment (typical of breathing-induced motion) to more substantial, clinically relevant misalignments, in line with previously reported registration errors of up to 10mm for lung SBRT patients during CBCT-CT alignment~\cite{oechsner2016registration}.

\subsection{Experimental Details}
The models were trained on a Ubuntu server using NVIDIA RTX A6000 graphics cards. Due to the large data size and 48 GB VRAM memory limit, CBCTLiTS, pancreas data and SynthRad volumes were isotropically downscaled by a factor of two~\cite{tschuchnig2024multi}. The dataset was split into training, validation, and testing subsets using a ratio of $0.7$ (training), $0.2$ (validation), and $0.1$ (testing). All experiments were conducted and evaluated using four different random splits to ensure stable and robust results. To enable comparability, the same data splits and random CT misalignments were applied across all model configurations. All model variants were trained for $100$ epochs using the Adam optimizer, with a learning rate of $1 \cdot 10^{-4}$, a weight decay of $1 \cdot 10^{-5}$ for regularization, and gradient accumulation over $8$ steps (effective batch size $8$) with a per-step batch size of $1$.

\section{Results}

\begin{figure}
\includegraphics[width=\textwidth]{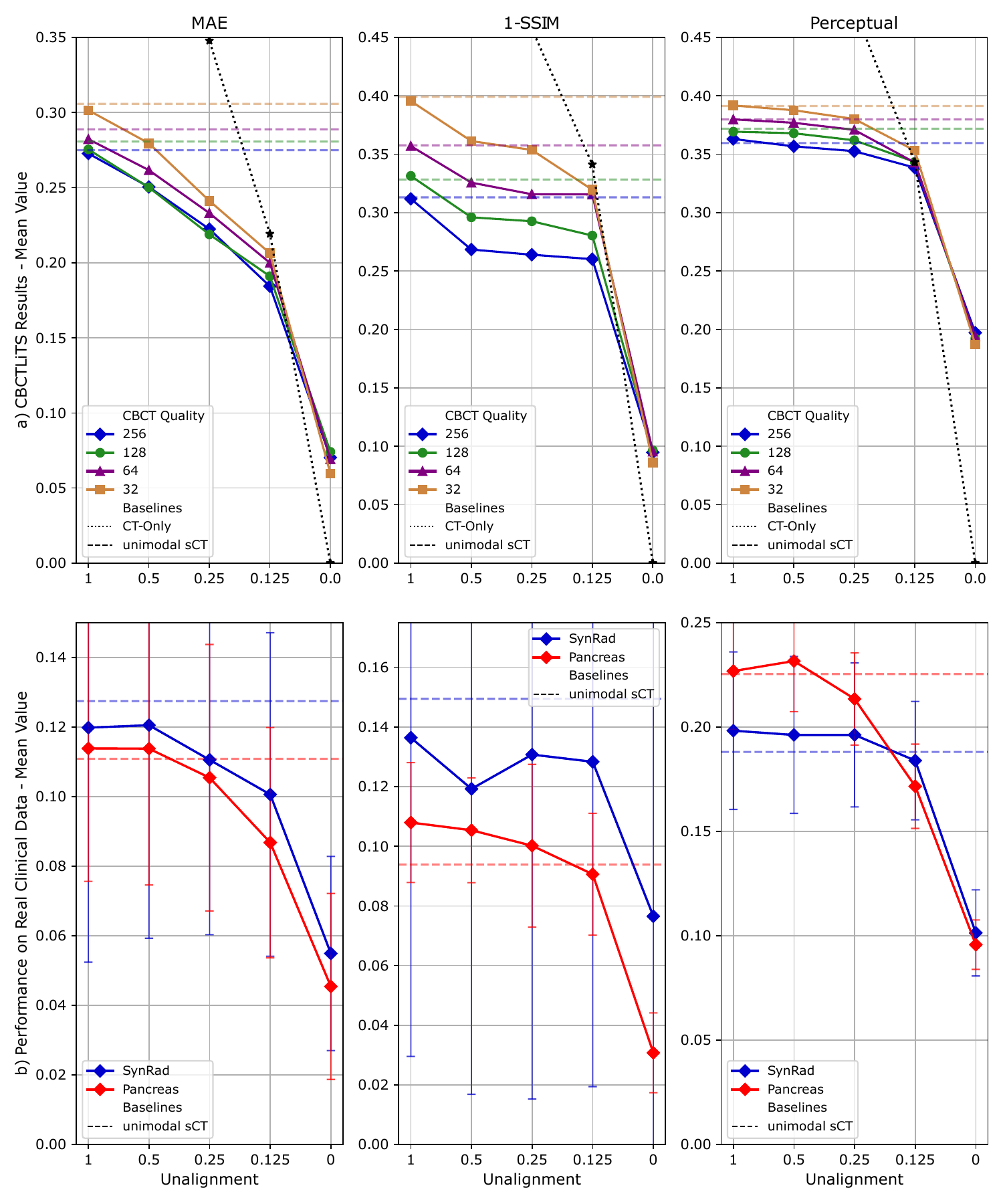}
\caption{Multimodal sCT results based on three metrics: MAE, \(\mathrm{1\text{-}SSIM}\) and perceptual dissimilarity. The x-axis indicates alignment level, from maximal unaligned to perfect alignment. The y-axis shows the mean value across four independent experiments. Solid lines represent our proposed multimodal method, with CBCT quality encoded by color. Dashed and dotted lines denote unimodal and CT-only baselines. CBCTLiTS results are shown in row a, pancreas and SynthRad results are shown in row b. Row b further displays uncertainty through error bars based on the corresponding standard deviation.} \label{Fig::Litsperceptual}
\end{figure}

Fig.~\ref{Fig::Litsperceptual}a presents the results of the proposed model alongside baseline methods on the CBCTLiTS dataset. Three evaluation metrics are shown: mean absolute error (MAE), \(\mathrm{1\text{-}SSIM}\), and the perceptual loss, each averaged over four independent experimental runs. To enhance interpretability, we report \(\mathrm{1\text{-}SSIM}\) instead of SSIM itself, ensuring that for all metrics, lower values indicate better performance.
The x-axis indicates the degree of spatial misalignment, ranging from maximal unalignment to perfect alignment (in the synthetic CBCTLiTS setup). Each subplot contains line plots for four CBCT quality levels, with color encoded CBCT quality, where blue ($\alpha_{np}=256$) corresponds to high-quality CBCT and orange ($\alpha_{np}=32$) to low-quality CBCT. Solid lines denote our multimodal model, dashed lines represent unimodal CBCT-only baselines, and dotted lines indicate CT-only baseline. For a complete overview of all CBCTLiTS results, including means and standard deviations across all experimental setups and baselines, see Table~\ref{tab:cbct_lits_results}.

Fig.~\ref{Fig::Litsperceptual}b shows corresponding results for the pancreas and SynthRad datasets. Here, SynthRad is plotted in blue and pancreas in red. Since CT-only reconstructions are infeasible in practical (real-world) settings, due to the absence of a sCT ground truth, they are omitted for these datasets. Additionally, uncertainty is visualized via error bars representing the standard deviation across the four runs. A full breakdown of results for the pancreas and SynthRad datasets can be found in Table~\ref{tab:cbct_pancreas_results} and Table~\ref{tab:cbct_synthrad_results}, respectively.

On the CBCTLiTS dataset (Fig.~\ref{Fig::Litsperceptual}a), our multimodal approach consistently outperforms both unimodal CBCT-only and CT-only baselines across all three metrics: MAE, \(\mathrm{1\text{-}SSIM}\), and perceptual dissimilarity. Performance gains are most pronounced in low-quality CBCT scenarios. For the perceptual loss, our model surpasses the CT-only baseline in nearly all conditions—except when CBCT quality is extremely low but CBCT-CT alignment is almost perfect.

\begin{table}[ht]
\centering
\caption{Evaluation Results (mean $\pm$ std) on the CBCTLiTS Dataset.}
\label{tab:cbct_lits_results}
\resizebox{\textwidth}{!}{%
\begin{tabular}{c | c | c | c | c | c | c | c}
\hline
$\alpha_{np}$ & $\alpha_a$ & MAE & CT-only & \(\mathrm{1\text{-}SSIM}\) & CT-only & Perceptual & CT-only \\
\hline
32 & & 0.306 $\pm$ 0.128 & & 0.399 $\pm$ 0.064 & & 0.391 $\pm$ 0.056 & \\
32 & 1 & 0.302 $\pm$ 0.116 & 0.764 $\pm$ 0.216 & 0.396 $\pm$ 0.070 & 0.668 $\pm$ 0.097 & 0.392 $\pm$ 0.054 & 0.666 $\pm$ 0.056 \\
32 & 0.5 & 0.280 $\pm$ 0.101 & 0.525 $\pm$ 0.124 & 0.361 $\pm$ 0.066 & 0.561 $\pm$ 0.079 & 0.388 $\pm$ 0.055 & 0.594 $\pm$ 0.061 \\
32 & 0.25 & 0.241 $\pm$ 0.075 & 0.348 $\pm$ 0.076 & 0.354 $\pm$ 0.075 & 0.458 $\pm$ 0.074 & 0.380 $\pm$ 0.054 & 0.475 $\pm$ 0.063 \\
32 & 0.125 & 0.206 $\pm$ 0.057 & 0.219 $\pm$ 0.050 & 0.320 $\pm$ 0.066 & 0.341 $\pm$ 0.066 & 0.353 $\pm$ 0.055 & 0.343 $\pm$ 0.057 \\
32 & 0 & 0.060 $\pm$ 0.020 & 0.000 $\pm$ 0.000 & 0.086 $\pm$ 0.044 & 0.000 $\pm$ 0.000 & 0.187 $\pm$ 0.019 & 0.000 $\pm$ 0.000 \\
\hline

64 & & 0.289 $\pm$ 0.122 & & 0.358 $\pm$ 0.061 & & 0.380 $\pm$ 0.057 & \\
64 & 1 & 0.282 $\pm$ 0.112 & 0.764 $\pm$ 0.216 & 0.357 $\pm$ 0.060 & 0.668 $\pm$ 0.097 & 0.380 $\pm$ 0.055 & 0.666 $\pm$ 0.056 \\
64 & 0.5 & 0.262 $\pm$ 0.093 & 0.525 $\pm$ 0.124 & 0.326 $\pm$ 0.057 & 0.561 $\pm$ 0.079 & 0.377 $\pm$ 0.056 & 0.594 $\pm$ 0.061 \\
64 & 0.25 & 0.233 $\pm$ 0.082 & 0.348 $\pm$ 0.076 & 0.316 $\pm$ 0.074 & 0.458 $\pm$ 0.074 & 0.371 $\pm$ 0.056 & 0.475 $\pm$ 0.063 \\
64 & 0.125 & 0.200 $\pm$ 0.062 & 0.219 $\pm$ 0.050 & 0.316 $\pm$ 0.073 & 0.341 $\pm$ 0.066 & 0.343 $\pm$ 0.054 & 0.343 $\pm$ 0.057 \\
64 & 0 & 0.069 $\pm$ 0.016 & 0.000 $\pm$ 0.000 & 0.096 $\pm$ 0.049 & 0.000 $\pm$ 0.000 & 0.194 $\pm$ 0.017 & 0.000 $\pm$ 0.000 \\
\hline

128 & & 0.281 $\pm$ 0.122 & & 0.328 $\pm$ 0.068 & & 0.372 $\pm$ 0.059 & \\
128 & 1 & 0.275 $\pm$ 0.113 & 0.764 $\pm$ 0.216 & 0.331 $\pm$ 0.072 & 0.668 $\pm$ 0.097 & 0.369 $\pm$ 0.055 & 0.666 $\pm$ 0.056 \\
128 & 0.5 & 0.250 $\pm$ 0.102 & 0.525 $\pm$ 0.124 & 0.296 $\pm$ 0.070 & 0.561 $\pm$ 0.079 & 0.368 $\pm$ 0.057 & 0.594 $\pm$ 0.061 \\
128 & 0.25 & 0.219 $\pm$ 0.087 & 0.348 $\pm$ 0.076 & 0.293 $\pm$ 0.069 & 0.458 $\pm$ 0.074 & 0.362 $\pm$ 0.055 & 0.475 $\pm$ 0.063 \\
128 & 0.125 & 0.191 $\pm$ 0.063 & 0.219 $\pm$ 0.050 & 0.280 $\pm$ 0.076 & 0.341 $\pm$ 0.066 & 0.343 $\pm$ 0.051 & 0.343 $\pm$ 0.057 \\
128 & 0 & 0.074 $\pm$ 0.021 & 0.000 $\pm$ 0.000 & 0.097 $\pm$ 0.048 & 0.000 $\pm$ 0.000 & 0.192 $\pm$ 0.016 & 0.000 $\pm$ 0.000 \\
\hline

256 & & 0.275 $\pm$ 0.126 & & 0.313 $\pm$ 0.072 & & 0.360 $\pm$ 0.062 & \\
256 & 1 & 0.273 $\pm$ 0.118 & 0.764 $\pm$ 0.216 & 0.312 $\pm$ 0.068 & 0.668 $\pm$ 0.097 & 0.363 $\pm$ 0.060 & 0.666 $\pm$ 0.056 \\
256 & 0.5 & 0.250 $\pm$ 0.097 & 0.525 $\pm$ 0.124 & 0.268 $\pm$ 0.069 & 0.561 $\pm$ 0.079 & 0.357 $\pm$ 0.059 & 0.594 $\pm$ 0.061 \\
256 & 0.25 & 0.222 $\pm$ 0.087 & 0.348 $\pm$ 0.076 & 0.264 $\pm$ 0.068 & 0.458 $\pm$ 0.074 & 0.353 $\pm$ 0.057 & 0.475 $\pm$ 0.063 \\
256 & 0.125 & 0.184 $\pm$ 0.063 & 0.219 $\pm$ 0.050 & 0.260 $\pm$ 0.070 & 0.341 $\pm$ 0.066 & 0.339 $\pm$ 0.050 & 0.343 $\pm$ 0.057 \\
256 & 0 & 0.070 $\pm$ 0.023 & 0.000 $\pm$ 0.000 & 0.095 $\pm$ 0.049 & 0.000 $\pm$ 0.000 & 0.197 $\pm$ 0.014 & 0.000 $\pm$ 0.000 \\
\hline
\end{tabular}
}
\end{table}

\begin{table}[ht]
\centering
\caption{Evaluation Results (mean $\pm$ std) on the Pancreas Dataset.}
\label{tab:cbct_pancreas_results}
\resizebox{\textwidth}{!}{%
\begin{tabular}{c | c | c | c | c | c | c | c}
\hline
$\alpha_a$ & MAE & CT-only & \(\mathrm{1\text{-}SSIM}\) & CT-only & Perceptual & CT-only \\
\hline
& 0.111 $\pm$ 0.037 & & 0.094 $\pm$ 0.026 & & 0.225 $\pm$ 0.025 & \\
1 & 0.114 $\pm$ 0.038 & 0.243 $\pm$ 0.070 & 0.108 $\pm$ 0.020 & 0.169 $\pm$ 0.049 & 0.227 $\pm$ 0.028 & 0.350 $\pm$ 0.065 \\
0.5 & 0.114 $\pm$ 0.039 & 0.150 $\pm$ 0.044 & 0.105 $\pm$ 0.018 & 0.135 $\pm$ 0.038 & 0.232 $\pm$ 0.024 & 0.272 $\pm$ 0.054 \\
0.25 & 0.105 $\pm$ 0.038 & 0.088 $\pm$ 0.027 & 0.100 $\pm$ 0.027 & 0.101 $\pm$ 0.028 & 0.213 $\pm$ 0.022 & 0.195 $\pm$ 0.038 \\
0.125 & 0.087 $\pm$ 0.033 & 0.051 $\pm$ 0.016 & 0.091 $\pm$ 0.020 & 0.068 $\pm$ 0.022 & 0.172 $\pm$ 0.020 & 0.136 $\pm$ 0.027 \\
0 & 0.045 $\pm$ 0.027 & 0.000 $\pm$ 0.000 & 0.031 $\pm$ 0.013 & 0.000 $\pm$ 0.000 & 0.096 $\pm$ 0.012 & 0.000 $\pm$ 0.000 \\
\hline
\end{tabular}
}
\end{table}

\begin{table}[ht]
\centering
\caption{Evaluation Results (mean $\pm$ std) on the SynthRad Dataset.}
\label{tab:cbct_synthrad_results}
\resizebox{\textwidth}{!}{%
\begin{tabular}{c | c | c | c | c | c | c}
\hline
$\alpha_a$ & MAE & CT-only & \(\mathrm{1\text{-}SSIM}\) & CT-only & Perceptual & CT-only \\
\hline
& 0.127 $\pm$ 0.059 & & 0.149 $\pm$ 0.138 & & 0.188 $\pm$ 0.039 & \\
1 & 0.120 $\pm$ 0.067 & 0.304 $\pm$ 0.091 & 0.136 $\pm$ 0.107 & 0.285 $\pm$ 0.110 & 0.198 $\pm$ 0.038 & 0.420 $\pm$ 0.068 \\
0.5 & 0.121 $\pm$ 0.061 & 0.187 $\pm$ 0.053 & 0.119 $\pm$ 0.102 & 0.194 $\pm$ 0.071 & 0.196 $\pm$ 0.038 & 0.350 $\pm$ 0.056 \\
0.25 & 0.111 $\pm$ 0.050 & 0.114 $\pm$ 0.032 & 0.131 $\pm$ 0.116 & 0.131 $\pm$ 0.046 & 0.196 $\pm$ 0.035 & 0.257 $\pm$ 0.044 \\
0.125 & 0.101 $\pm$ 0.047 & 0.069 $\pm$ 0.020 & 0.128 $\pm$ 0.109 & 0.086 $\pm$ 0.032 & 0.184 $\pm$ 0.028 & 0.181 $\pm$ 0.034 \\
0 & 0.055 $\pm$ 0.028 & 0.000 $\pm$ 0.000 & 0.077 $\pm$ 0.111 & 0.000 $\pm$ 0.000 & 0.101 $\pm$ 0.021 & 0.000 $\pm$ 0.000 \\
\hline
\end{tabular}
}
\end{table}

Applied to the pancreas dataset our multimodal model underperforms in highly misaligned cases but yields improvements over unimodal CBCT-based sCT generation when alignment improves. Specifically, performance exceeds the unimodal baseline for MAE and perceptual loss at alignment levels $\alpha_a \leq 0.25$, and for \(\mathrm{1\text{-}SSIM}\) at $\alpha_a \leq 0.125$. Results on SynthRad closely mirror those observed on CBCTLiTS. Here, the multimodal model consistently achieves lower MAE and \(\mathrm{1\text{-}SSIM}\) scores compared to the unimodal baseline. Perceptual improvements are dependent on alignment, emerging in well aligned scenarios.

\begin{figure}[t]
\includegraphics[width=\textwidth]{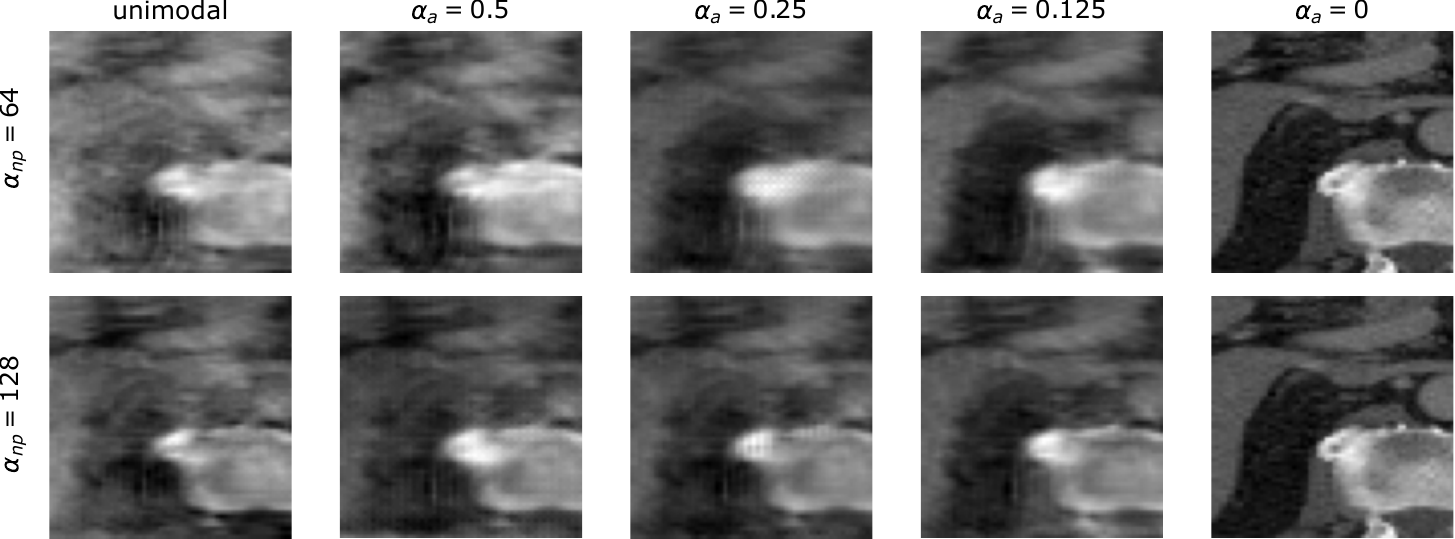}
\caption{
Qualitative comparison of sCT reconstructions on the CBCTLiTS dataset. Rows correspond to different CBCT quality levels (top: low, bottom: high). Columns compare methods: a unimodal CBCT baseline (left) and multimodal reconstructions conditioned on preoperative CTs with increasing alignment (right). Images are normalized for visualization and shown at increased zoom to better highlight anatomical structures.}
\label{Fig::qual}
\end{figure}

Fig.~\ref{Fig::qual} presents representative sCT slices shown at increased zoom to emphasize anatomical details across different CBCT qualities and alignment levels. The unimodal baseline and three multimodal reconstructions are shown. With higher CBCT quality and better alignment, reconstructions exhibit clearer anatomical structures and fewer artifacts.

\section{Discussion}
Our multimodal model consistently achieves lower MAE, higher SSIM, and improved perceptual similarity compared to unimodal sCT approaches, underscoring the value of fusing intraoperative CBCT with preoperative CT. These improvements generalize across all datasets, with two exceptions: slightly lower SSIM on the pancreas dataset and reduced perceptual performance in real-world cases with moderate to strong CBCT–CT misalignment. While the perceptual degradation remains modest, it appears consistently across all real-world datasets in cases of moderate to strong misalignment. This suggests that while multimodal fusion effectively reduces CBCT artifacts (as seen in lower MAE), it may slightly compromise high-frequency perceptual details when alignment is poor. Under high alignment conditions, however, the multimodal model consistently outperforms unimodal baselines across all evaluation metrics.

Using the perfectly aligned CBCT–CT pairs from the CBCTLiTS dataset, we show that the multimodal model does not simply replicate the preoperative CT. Instead, it effectively fuses structural detail from the CT and intraoperative CBCT. This is evidenced by the model outperforming both CT-only and CBCT-only baselines simultaneously across all alignment levels and CBCT quality settings. In partially misaligned cases, we hypothesize that the model performs implicit spatial registration, allowing relevant anatomical features from both modalities to be aligned and fused during training, improving reconstruction quality.

Our investigation into CBCT quality ($\alpha_{np}$) and CBCT–CT alignment ($\alpha_a$) demonstrates that both factors critically impact sCT reconstruction quality. Across all metrics, we consistently observe improved performance with increasing CBCT quality, suggesting that higher-quality intraoperative scans benefit reconstruction irrespective of the input modality (unimodal or multimodal). Likewise, better alignment leads to continuous gains across most experimental configurations, without evidence of saturation. This suggests that further improvements, e.g., via manual pre-registration or integrated mechanisms such as spatial transformer networks~\cite{jaderberg2015spatial}, may yield additional performance boosts. These findings underline the dual importance of enhancing CBCT quality and optimizing alignment to maximize sCT reconstruction fidelity.

On the real-world pancreas dataset, the multimodal model improves upon unimodal sCT in terms of MAE and perceptual scores for alignment levels  \(\alpha_a \leq 0.25\). However, SSIM performance degrades in all but the most well-aligned cases, often resulting in overly bright reconstructions with reduced structural contrast. We attribute this to the dataset’s limited size and substantial anatomical variability, which likely hinder the model’s ability to learn consistent spatial and textural representations. In contrast, results on the real-world SynthRad dataset closely mirror those of the synthetic CBCTLiTS experiments, particularly for MAE and SSIM. Perceptual scores on SynthRad follow similar trends but are generally slightly worse than those observed in the synthetic setting. These findings suggest that the SSIM degradation in the pancreas dataset is not due to a fundamental domain shift, but rather reflects data scarcity and greater anatomical complexity.

Finally, the CBCTLiTS dataset provides voxel-wise alignment between CBCT and CT, enabling ideal fusion conditions. This setup enables ideal conditions for both training and evaluation. In contrast, real-world datasets inherently exhibit spatial misalignment between CBCT and CT scans, making precise reconstruction more challenging, especially given the lack of a ground truth sCT. These observations emphasize the critical role of spatial alignment in training multimodal sCT models and support the utility of synthetic datasets with perfect correspondence as a valuable tool for developing and benchmarking such models.

\section{Conclusion}
We present a multimodal approach for sCT generation that fuses intraoperative CBCT with preoperative CT data in a 3D framework. Our method extends prior 2D models and is systematically evaluated across both synthetic and real-world datasets. Results demonstrate that multimodal sCT consistently outperforms unimodal baselines in terms of MAE and SSIM, especially under conditions of high CBCT quality and good alignment.

Our analysis highlights the critical roles of CBCT quality and CBCT–CT alignment: both factors significantly influence reconstruction fidelity, and we observe no performance plateau, indicating room for further improvement. Notably, even in the presence of substantial misalignment, our model yields superior reconstructions compared to unimodal approaches, although perceptual scores degrade in such cases.

Experiments on the real-world pancreas and SynthRad datasets confirm the transferability of insights from the synthetic CBCTLiTS benchmark. While SSIM degrades in the low-data, high-variability pancreas setting, SynthRad matches synthetic results closely, reinforcing our method’s generalizability.

Finally, we underscore the value of synthetic datasets with perfect alignment as development tools for multimodal fusion, and call attention to the need for alignment-aware architectures and training strategies. Future work should focus on improved perceptual fidelity, implicit registration mechanisms, and clinical validation to position multimodal sCT as a practical tool for intraoperative imaging.

\begin{credits}
\subsubsection{\ackname} 
Blinded

\subsubsection{\discintname}
Not applicable

\end{credits}
%
%
%
\bibliographystyle{splncs04}
\bibliography{mybibliography}
\end{document}